# Interfacial Coupling Controls Molecular Epitaxy of HMTP on Graphene/SiC

Devanshu Varshney,[1&] Pavel Procházka,[2&] Veronika Stará,[2] Mykhailo Shestopalov,[3] Jan Kunc,[3*] Jiří Novák,[1*] Jan Čechal[2,4*]

[1]Department of Condensed Matter Physics, Faculty of Science, Masaryk University, Kotlářská 2, 61137 Brno, Czech Republic

[2]CEITEC - Central European Institute of Technology, Brno University of Technology, Purkyňova 123, 612 00 Brno, Czech Republic.

[3] Charles University, Faculty of Mathematics and Physics, Institute of Physics, Ke Karlovu 5, 121 16, Prague 2, Czech   Republic

[4] Institute of Physical Engineering, Brno University of Technology, Technická 2896/2, 616 69 Brno, Czech Republic.

[&]These authors contributed equally.

**Abstract**

Epitaxial growth critically influences structural and electronic properties of organic semiconductors. Graphene serves as a prominent van der Waals template for molecular self-assembly; however, graphene on SiC is intrinsically heterogeneous, with decoupled monolayer graphene coexisting with residuals of a covalently bound buffer layer, which may affect molecular ordering. Here, we track the ordering of the molecular donor, 2,3,6,7,10,11-hexamethoxytriphenylene (HMTP), from the first layer to thin films, combining low-energy electron microscopy and diffraction with X-ray diffraction. HMTP forms highly ordered epitaxial layers on single-layer graphene, whereas growth on the buffer layer initiates as amorphous and evolves into polycrystalline films with weak orientation with respect to the substrate. Crucially, hydrogen intercalation decouples the buffer layer, converting it into quasi-freestanding monolayer graphene and restoring epitaxial growth. These findings demonstrate that interfacial coupling governs molecular epitaxy on graphene/SiC, and interface engineering via hydrogen intercalation provides a scalable route to control organic thin-film crystallinity on graphene.

**Keywords**

## Introduction

Crystallinity and orientation of organic molecular thin films significantly influence their charge transport and optical properties, thereby directly affecting device performance.[1–3] The growth of organic films is largely determined by the substrate, which can promote epitaxy, induce disorder, or stabilize additional, surface-induced molecular arrangements.[4–6] Among possible substrates, graphene, a two-dimensional material with atomically flat and well-defined surface, provides a robust van der Waals template for assembly of planar π-conjugated molecules due to its π-electron system,[7–9] making graphene an essential component for organic–inorganic van der Waals hybrid materials.[7–11]

Epitaxial graphene grown on SiC is particularly promising due to its technological relevance and wafer-scale availability.[12,13] However, graphene on SiC is intrinsically heterogeneous: decoupled single-layer graphene (SLG, Figure 1b) coexists with a carbon buffer layer that is partially covalently bonded to the SiC substrate (Figure 1a).[13] Although the buffer layer retains a graphene-like lattice, the bonding of substrate Si atoms to carbon atoms in the buffer layer disrupts π-conjugation. This results in markedly different local structural and electronic properties that make the buffer layer intrinsically heterogeneous on a nanoscale.[13,14] While previous studies have reported long-range ordered growth of organic semiconductor monolayers on graphene[8,14–22] and disordered growth on the buffer layer,[14,17,23] a direct experimental link between interface-specific monolayer ordering and the macroscopic organic-film crystallinity is still missing. It remains unclear whether the buffer layer merely disrupts local order or fundamentally prevents the formation of crystalline and epitaxial thin films.

Here, we bridge the gap between the local interfacial structure and crystalline order of the thin film by combining low-energy electron microscopy and diffraction (LEEM/LEED),[24,25] which probe molecular order during the initial growth phase, with X-ray diffraction (XRD), which characterizes the crystallinity of thin films. Using 2,3,6,7,10,11-hexamethoxytriphenylene (HMTP, Figure 1d)[26,27] as a prototypical planar organic semiconductor, we show that highly ordered, epitaxial HMTP layers form selectively on graphene. In contrast, the growth on the buffer layer starts as laterally disordered (amorphous) but evolves crystalline order with small grains of random in-plane orientation as the film thickens grows. However, the coplanar molecular orientation with the substrate is preserved throughout growth. Crucially, hydrogen intercalation decouples the buffer layer from the SiC substrate, converting it into quasi-freestanding SLG (Figure 1c),[13] and restores epitaxial growth. This directly links interfacial coupling to thin-film crystallinity and demonstrates that interface engineering via hydrogen intercalation provides a scalable route to wafer-scale control of organic molecular epitaxy on graphene.

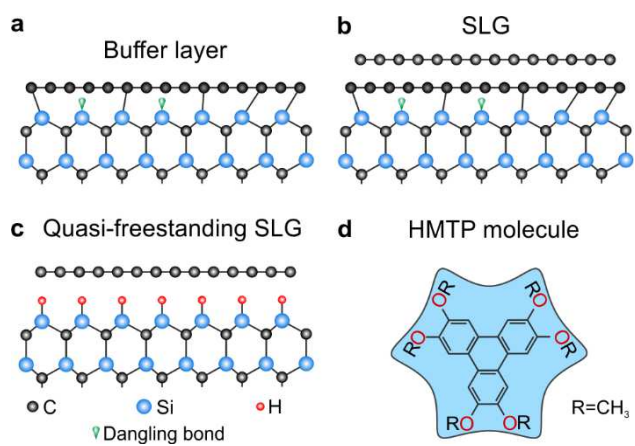

**Figure 1**: Schematic overview of the substrates and the molecule considered in this work. (a) SiC surface with the carbon buffer layer, which retains a graphene-like structure but remains partially covalently bonded to substrate Si atoms. (b) Single-layer graphene (SLG) grown on top of the buffer layer. (c) Hydrogen intercalation of the buffer layer passivates the Si–C bonds, converting it into quasi-freestanding SLG. (d) Molecular structure of 2,3,6,7,10,11-hexamethoxytriphenylene (HMTP) molecule.

**Results and Discussion**

First, we present the XRD results of HMTP thin films deposited on two different substrates: single-layer graphene (SLG) and a buffer layer. These substrates are prepared via thermal annealing of 6H-SiC and characterized by Raman spectroscopy[28] as described in Supporting Information, Section 1. Initially, the annealing results in Si desorption leaving carbon atoms that form a precursor buffer layer with a graphene-like structure, which remains covalently bound to the substrate Si atoms.[13] Continued Si desorption leads to the formation of a new buffer layer, allowing the decoupling of the initial one as graphene. Due to the intrinsic heterogeneity of the surface, it is not possible to prepare pure samples consisting entirely of the buffer layer or SLG; samples always contain either traces of graphene on a nominal buffer layer or residual buffer layer on a nominal SLG sample. The XRD results for samples with coexisting buffer and SLG at a ratio of approximately 2:1 are provided in Supporting Information, Section 2. We then take advantage of the coexistence of the buffer layer and SLG to perform spatially resolved LEEM/LEED studies under identical deposition conditions, revealing distinct growth on both parts of the sample. Finally, we introduce hydrogen intercalation to break the buffer-layer bonds to the substrate, transforming it into quasi-freestanding SLG and restoring the epitaxial growth of HMTP. Here, epitaxial growth refers to the formation of a long-range ordered molecular film that follows the orientation of the underlying substrate lattice.

**X-ray diffraction**. Crystallographic texture and crystal quality of HMTP thin films with a thickness of 28 – 33 nm were characterized by X-ray diffraction pole figures, azimuthal scans, symmetric (ω/2θ) scans, and rocking (ω) scans for samples on both kinds of substrates. While the pole figures and azimuthal scans provide an overview of the film texture, the azimuthal and rocking scans provide more quantitative information about the crystal mosaicity within the sample plane (referred to as the in-plane direction) and in the direction perpendicular to the surface plane (referred to as the out-of-plane direction), respectively. The symmetric scan provides information about the crystal quality along the out-of-plane direction.

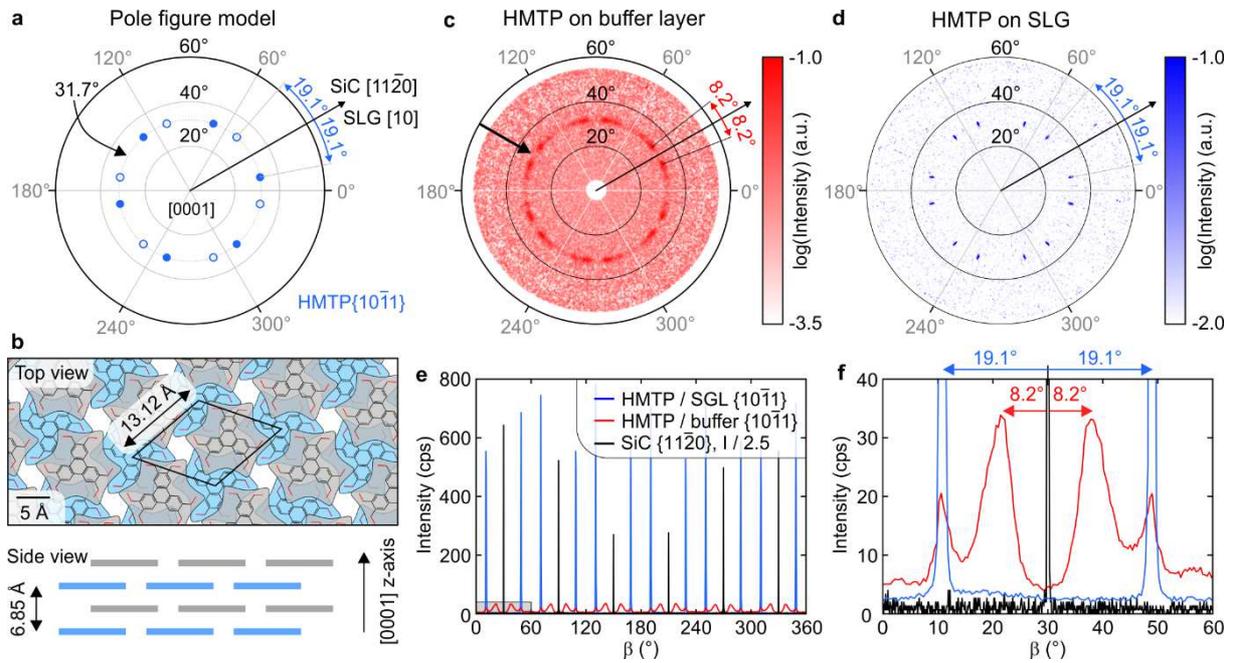

**Figure 2:** Pole figures, structural model, and azimuthal scans of HMTP layers on the buffer layer and SLG. (a) Schematic pole figure model for the HMTP $\{10\bar{1}1\}$ reflections. The poles are located at a polar angle of 31.7° relative to HMTP [0001] direction perpendicular to the sample surface, corresponding to the inclination of the $\{10\bar{1}1\}$ planes, and appear in two 6-fold sets due to the presence of two rotational domains. (b) Real-space structural model of the bulk HMTP crystal, illustrating the in-plane molecular arrangement and interlayer stacking geometry. (c, d) XRD pole figures for HMTP $\{10\bar{1}1\}$ reflections measured on the buffer layer (c) and on SLG (d). The ring-shaped band of enhanced intensity (indicated by the arrow in (c)) and sharp spots in (d) are both centered at a polar angle of 31.7° with respect to the HMTP [0001] direction. (e) Azimuthal (β) scans of the HMTP $\{10\bar{1}1\}$ reflections at the polar angle 31.7° for SLG (blue) and the buffer layer (red), together with SiC substrate $\{11\bar{2}0\}$ in-plane reflections (black). The azimuthal angle β was aligned using the SiC $\{11\bar{2}0\}$ reflections as reference. The SiC intensity is downscaled by a factor of 2.5 for clarity. (f) Enlarged view of the azimuthal scans shown in (e).

The pole figures for the HMTP $\{10\bar{1}1\}$ reflections of HMTP films (see schematics in Figure 2a) on both substrates are shown in Figure 2c and d. For the film grown on the buffer layer, we observe a ring-shaped band of enhanced intensity with a slight azimuthal modulation marked with an arrow in Figure 2c. The radial position of the band implies that the HMTP $\{10\bar{1}1\}$ poles are inclined by 31.7° with respect to the sample surface normal. As the ring-shaped band appears symmetrically with respect to the pole figure center, the HMTP $\{0001\}$ lattice planes are parallel to the sample surface. These results are consistent with the HMTP bulk structure,[27] which is placed on the sample surface, with molecules coplanar with the surface (Figure 2b). The weak but visible azimuthal intensity modulation indicates that the film is composed of many domains with weak in-plane orientation order. We will further discuss this in the context of the azimuthal scans. On the other hand, the pole figure measured on the SLG sample (Figure 2d) reveals two sets of six sharp spots corresponding to two mirror domains of the 6-fold-symmetric molecular layer on the sample surface. Contrary to the buffer layer, the HMTP domains on SLG have almost perfect orientation within the sample surface plane. The radial position of all diffraction spots is 31.7°, i.e., the same as that of the intense band of the buffer layer. Thus, HMTP films grow with $\{0001\}$ lattice planes parallel to the sample surface on both substrates, but only on SLG do they exhibit strong in-plane preferential orientation.

The detailed analysis of the in-plane distribution of HMTP domains for both substrate types was done using the azimuthal scans shown in Figures 2e and f. We have probed the HMTP $\{10\bar{1}1\}$ poles radially inclined by 31.7° in the pole figures. To provide a reference to the orientation of the substrate lattice, we have measured azimuthal scans for the SiC

{11$\bar{2}$0} reflections. As the SiC [11$\bar{2}$0] crystallographic direction is parallel to the [10] direction (note that we use 2D notation for graphene) of both the buffer layer and SLG, the azimuthal positions of SiC {11$\bar{2}$0} reflections coincide with those of {10} reflections for both the buffer layer and SLG. The angular positions of all peaks in the azimuthal scan match those of the pole figures and exhibit the same symmetry. For the SLG sample, the two sets of HMTP {10$\bar{1}$1} reflections are azimuthally rotated by ±19.1° out of SiC {11$\bar{2}$0} reflections. Therefore, the in-plane ⟨10$\bar{1}$1⟩ lattice directions of the two HMTP domains are rotated by ±19.1° with respect to the associated SLG ⟨10⟩ directions. The rotation angle between the HMTP lattice and SLG is consistent with our LEED measurements below, and with the HMTP monolayer on graphene/Ir(111)[22] and graphene/Ni(111)[29]. The FWHM of the peaks is 0.5°, which is, however, comparable to the resolution of the experimental setup. This confirms superior in-plane orientation ordering of the HMTP lattice on the SLG sample.

In contrast to SLG, the maxima of HMTP {10$\bar{1}$1} reflections are ~20 times less intense on the buffer layer and substantially broader. Moreover, we observe additional HMTP-related peaks, as detailed in Figure 2f. The additional peaks have a rotation angle of ±8.2° with respect to the nearest SiC {11$\bar{2}$0} reflection, and the FWHM of ~6°. Also, the peaks that appear at the same angular rotations as in the SLG sample (±19.1°) have a FWHM of 2°, i.e., they are much broader than in the SLG sample. The larger FWHM of the peaks indicates greater in-plane mosaicity of the HMTP domains on the buffer layer, i.e., a greater spread of possible HMTP grain orientations with respect to the substrate. Moreover, the azimuthal scan of HMTP on the buffer layer shows a 4 times higher background than on SLG, indicating a significant fraction of crystal domains with a random in-plane lattice orientation. The integrated area of the azimuthal scans measured on the buffer layer (Figures 2e and f) is approximately 70 % of that measured on SLG; since the intensity scales with the square of the domain/crystallite size within kinematic approximation and the smaller size of islands on the buffer layer (see below), we conclude that the vast majority of the HMTP layer is crystalline. In summary, we observe a strong and unique in-plane orientational order of HMTP domains on SLG. On the buffer layer, the overall crystallinity is comparable with SLG but with random in-plane orientation of the crystallites showing a weak orientational preference.

To address the crystal ordering in the out-of-plane direction, we have probed diffraction points with pure out-of-plane components via symmetric coplanar XRD scans (ω/2θ) and rocking scans (ω) through observed reflections. The symmetric scans probe crystal planes parallel to the sample surface. The scans through the HMTP 0002 reflection are shown in Figure 3a; the data for the SiC 000$l$ reflection series are given in Supporting Information, Section 3. The peak intensity is higher for HMTP on SLG than for HMTP on the buffer layer, attesting to higher crystal quality, i.e., fewer defects and larger crystal coherence length, for the SLG sample along the out-of-plane direction. Additionally, the 0002 reflection is surrounded by Laue (i.e., finite crystal thickness) oscillations for SLG but not for the buffer layer. This reflects the high crystal quality of the film and signals that the different HMTP domains have approximately the same thickness for the SLG. The out-of-plane mosaicity is addressed by rocking scans through the HMTP 0002 reflection shown in Figure 3b. The FWHM of the rocking curves for the buffer layer is slightly larger than that for SLG, but both are below 0.01°, which is close to the resolution limit of the experimental setup. This represents extraordinarily low mosaicity for the organic semiconductor thin films comparable to diindenoperylene on $SiO_2$,[30] but two orders of magnitude lower than that of π-stacked pentacene on graphene[31]. The very low out-of-plane mosaicity of both samples indicates that HMTP average molecular planes in all crystal grains/domains are parallel with the sample surface. Still, the thickness of the crystal domains is more uniform on SLG.

To sum up, the XRD data are consistent with the HMTP bulk structure[27], placed on the sample surface, such that the molecules are coplanar with the surface. The HMTP layer grows epitaxially on SLG, forming two mirror domains. On the buffer layer, the molecular plane remains coplanar with the sample surface, but most domains are randomly in-plane oriented with only a weak preferential orientation.

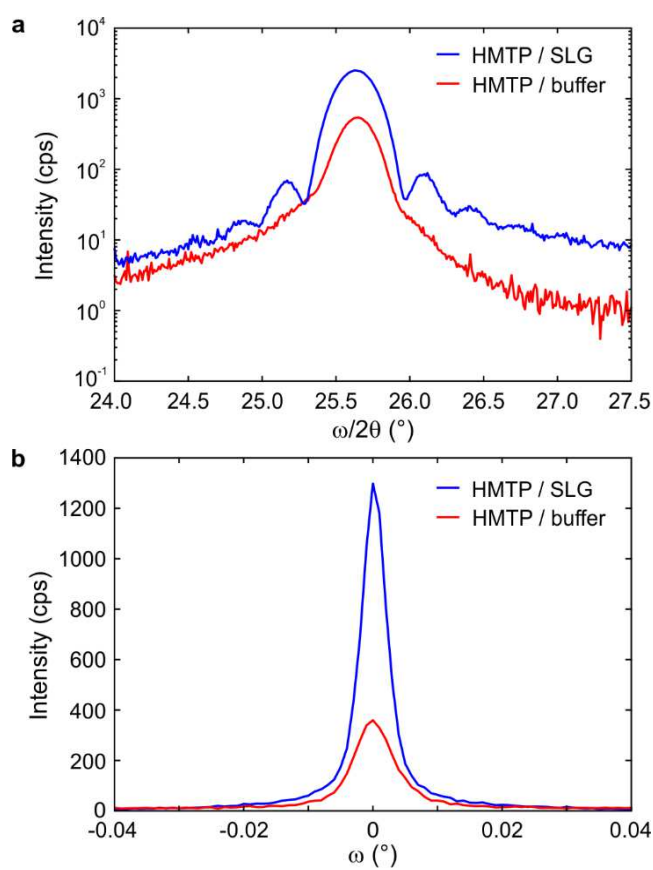

**Figure 3**: Symmetric and rocking scans for films grown on SLG (blue) and buffer layer (red). (a) Symmetric ω/2θ scan through the HMTP 0002 reflections. The Laue oscillations next to the main peak are clearly visible on the HMTP on SLG. (b) Rocking scan through the HMTP 0002 reflections.

**Atomic Force Microscopy**. The morphology of the films probed by AFM is shown in Figure 4. On the buffer layer, HMTP forms small, laterally separated islands of ~0.1 μm in size (Figure 4a), featuring well-developed facets often inclined at 60° or 120°, thus reflecting the in-plane hexagonal symmetry of the HMTP crystals, reminiscent of the HMTP crystals found in the powder[32]. However, the in-plane rotation of the islands is random. On the other hand, HMTP on SLG (Figure 4b) forms a flat, percolated layer. The line profile shows that the cracks are either shallow or very narrow, so their true depth cannot be measured with the AFM. The uniformity of the film on SLG is also reflected in the lower root-mean-square roughness of ~4 nm, compared to ~9 nm on the buffer layer. The contrast between the morphologies of HMTP on the buffer layer and SLG is also clearly visible on samples where both surface types coexist, as shown in Figure S6 in the Supporting Information. There, a flat HMTP layer covers SLG areas, while small islands cover the buffer layer. The observed morphology fully corroborates the XRD conclusions about the epitaxially grown islands of uniform thickness on SLG and the randomly oriented crystals of varying height on the buffer layer.

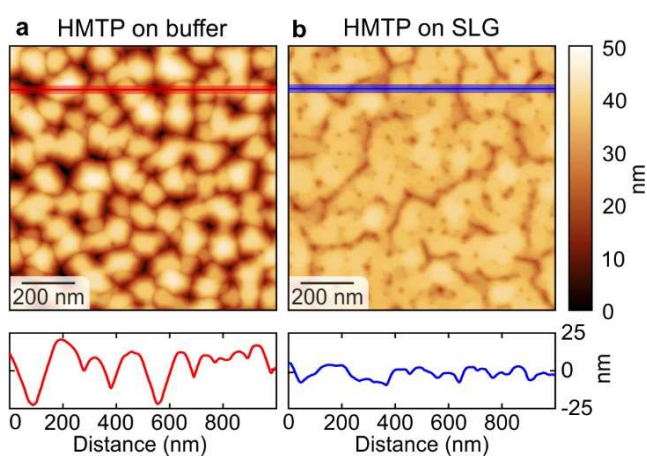

**Figure 4:** AFM images of the HMTP film on (a) the buffer layer and (b) SLG. The large-scale images are given in Supporting Information, Section 4.

**Low-energy electron microscopy.** To determine whether the structural disorder observed on the buffer layer originates at the molecule–substrate interface or develops later during film growth, we performed real-time LEEM measurements during HMTP deposition on a sample containing both buffer layer and SLG. This configuration allows for a direct, side-by-side comparison of the growth on the two surfaces under identical deposition conditions. In the bright-field image in Figure 5a, SLG regions appear brighter than the buffer layer areas, enabling straightforward distinction of both substrates.

A time-resolved sequence of bright-field images recorded during deposition is shown in Figure 5b; the real-time video is provided as Supporting Video SV1. After 100 s of HMTP exposure, a pronounced change in contrast is observed exclusively on SLG. Compact, ordered HMTP islands nucleate and laterally expand through the SLG area (highlighted by the purple box). In contrast, on the neighboring buffer layer region (white box), no such islands form. This behavior is quantified in Figure 5c, which displays the normalized bright-field intensity extracted from the two marked regions. On SLG, the intensity first decreases slowly during the initial 100 s of deposition, indicating an increase in the density of HMTP molecules at submonolayer coverages. This is followed by a sharp drop in intensity caused by the formation of an ordered HMTP layer. By contrast, the buffer layer exhibits only a gradual, spatially uniform, monotonic decrease in intensity, consistent with the accumulation of molecular species rather than the nucleation and growth of an ordered overlayer.

To confirm that the ordered HMTP overlayer forms exclusively on SLG, we thoroughly analyzed the sample after 400 s of deposition, i.e., once the compact HMTP islands fully covered all SLG regions by 1 monolayer (ML). The bright-field image in Figure 5d marks two circular areas on SLG, from which the diffraction was measured. The corresponding single-domain diffraction patterns in Figure 5e reveal two distinct HMTP orientations on SLG. Aside from moiré spots arising from the relative alignment of the underlying SLG and SiC substrate (analyzed in detail in Supporting Information, Section 5), both patterns contain sharp, well-defined diffraction spots, demonstrating that the HMTP overlayer on SLG is highly ordered and forms well-oriented crystalline domains. Both HMTP superlattices are structurally equivalent and form commensurate structures with the graphene lattice. Their unit cell corresponds to a $2\sqrt{7} \times 2\sqrt{7}$ R19.1° superstructure,[22] which can be expressed in the matrix notation as $\begin{pmatrix} 4 & 2 \\ 6 & -4 \end{pmatrix}$, as illustrated by a model provided in the Supporting Information, Section 6. Here, the HMTP bulk unit cell is by 0.1 Å larger than the superstructure unit cell ($2\sqrt{7} \times 2\sqrt{7}$ R19.1°), i.e., the lattice mismatch is below 1 %. The diffraction spots associated with the two orientations are combined into the color-coded composite shown in Figure 5f, which is fully consistent with the large-area diffraction pattern shown in Figure S10a in the Supporting Information.

Next, we have employed the phase-imaging capability provided by dark-field imaging;[24] there, selecting a single diffraction spot with an aperture allows us to image only the associated structure. Dark-field images were acquired using the diffraction spots highlighted in Figure 5f, and their color-coded composition in Figure 5g maps the spatial distribution of the two HMTP domain orientations. Dark-field contrast unambiguously shows that the ordered HMTP phase forms exclusively on the SLG, with no crystalline diffraction or domain contrast detectable on the buffer layer. The corresponding individual dark-field images for each orientation are shown in Supporting Information, Section 7.

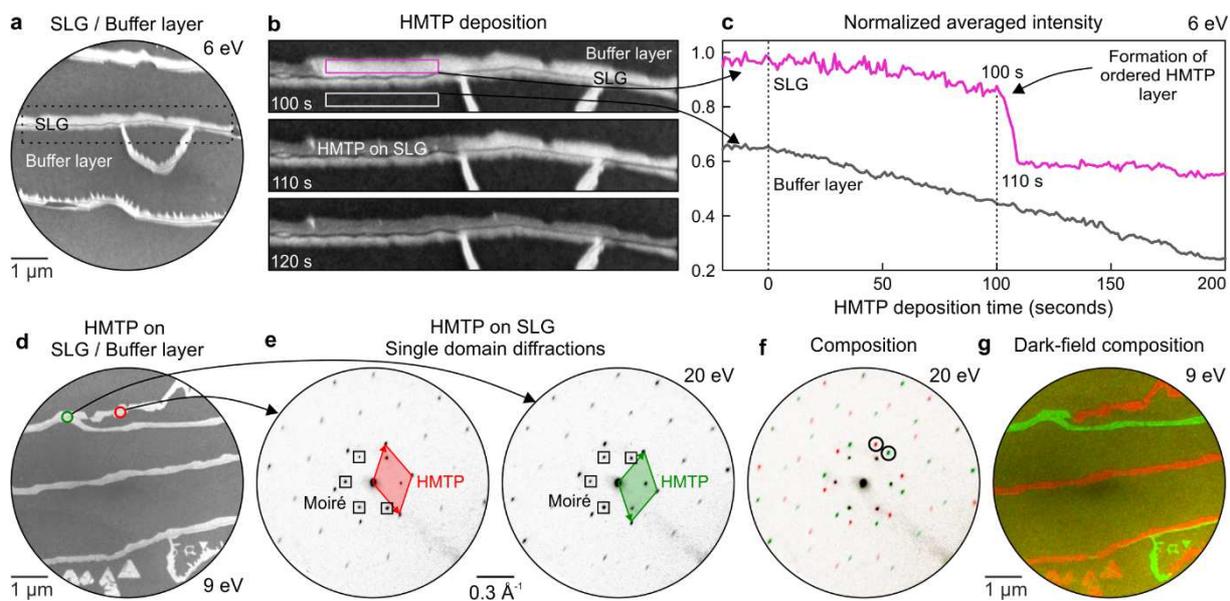

**Figure 5**: LEEM analysis of HMTP growth on SLG and the buffer layer. (a) Bright-field image showing SLG (bright) and buffer layer (dark) regions on SiC before HMTP deposition. (b) Time-lapse bright-field images recorded during HMTP deposition. Compact ordered HMTP islands nucleate and grow only on SLG (purple rectangle), whereas the buffer layer (white rectangle) exhibits only a gradual decrease in intensity. (c) Corresponding intensity evolution extracted from the regions highlighted in (b). (d) Bright-field image acquired at 9 eV after 400 s of HMTP deposition, marking two circular areas on SLG from which diffraction was collected. (e) Single-domain diffraction patterns from the two regions in (d), revealing two distinct HMTP orientations on SLG. Moiré spots, marked by black rectangles, arise from the relative alignment of the underlying SLG and SiC substrate. (f) Color-coded composite of the diffraction spots corresponding to the two HMTP orientations. (g) The dark-field composite image from the spots in (f) maps the spatial distribution of both HMTP orientations.

In contrast to SLG, no HMTP diffraction spots were detected on the buffer layer when the SLG regions were fully covered by a compact HMTP monolayer, apart from the substrate-related moiré pattern, as illustrated in the <span style="color:teal">Supporting Information, Section 8</span>. Additionally, a deposition of up to $10 \pm 2$ ML (~3.5 nm) did not produce any detectable HMTP diffraction on the buffer layer. Instead, the substrate moiré spots progressively weakened and became diffused, indicating the gradual accumulation of a poorly ordered HMTP layer. Overall, these observations show that HMTP initially grows amorphously on the buffer layer, and no crystalline order develops within the first layers. These results suggest that the polycrystalline nature of the layer observed by XRD develops during the growth of HMTP film.

**Discussion.** The results presented above show that HMTP grows epitaxially exclusively on SLG. The epitaxial growth of HMTP can be understood in terms of relatively strong physisorption, with significant orientational preference of triphenylene core for the graphene substrate via π–π interactions,[33] and relatively weak intermolecular interactions.[29] The epitaxial relationship of the first layer of molecules is not uncommon even for physisorbed molecules weakly interacting with substrate,[34] a phenomenon often referred to as van der Waals Epitaxy.[35,36] Molecule-surface interactions and substrate templating can induce the formation of new molecular phases with structures distinct from the bulk.[4] With increasing film thickness, the structure either smoothly approach the bulk lattice or, at a specific critical thickness, undergo a transition to a polycrystalline film.[37] In the case of HMTP on SLG, such a transition does not occur, as it grows in the bulk structure[27] adopted from the first layer.

In contrast with the SLG, initially disordered growth is observed on the buffer layer, which later evolves into the polycrystalline thin film in which all crystals are coplanar with the surface. The buffer layer is structurally similar to SLG but contains a significant number of $sp^3$-hybridized carbon atoms, which form three bonds within the graphene-like layer and one bond with the underlying Si. The surface of the buffer layer is thus both structurally and electronically non-uniform. Due to the structural similarity with graphene, the buffer layer is also referred to as "zero-layer graphene"[17] or "interfacial graphene"[14] or, if the lateral inhomogeneity is emphasized, as "SiC nanomesh"[23,38]. The origin of the small HMTP crystals with almost random in-plane orientation can be traced to the first layer, indicating a substrate effect. Previous studies show that copper and Cl-aluminum phthalocyanines form a highly ordered single-molecular array of nonintegrating, isolated molecules, whose positions are fully defined by the periodicity and symmetry of the buffer layer.[23,38] Conversely, pentacene grows in a disordered fashion,[17,23] with this disorder attributed to hindered surface diffusion due to the high density of trapping sites;[17] this initial disorder is then propagated into thicker layers. The evolution of the bright-field intensity in LEEM is consistent with limited diffusion of HMTP on the buffer layer suggesting that the scenario described for pentacene is also plausible for HMTP. The disorder is thus linked to $sp^3$ hybridization and resulting structural and electronic inhomogeneity.

If the $sp^3$ hybridization of carbon atoms is indeed responsible for the disordered growth, then breaking the bonding with the substrate Si atoms should render all the carbon atoms in the buffer layer $sp^2$ hybridized, thereby restoring the capability of the substrate to support epitaxial growth of HMTP. To test this, we have passivated the substrate Si with hydrogen, resulting in the formation of quasi-freestanding monolayer graphene. Indeed, on such a substrate, XRD shows epitaxial growth of HMTP.

**HMTP on quasi-freestanding SLG**. The HMTP layers grown on the quasi-freestanding SLG are very similar to those on SLG. The pole figure (Figure 6a) and azimuthal scans (Figure 6c) closely resemble those for SLG, indicating comparable in-plane orientational order. A symmetric coplanar XRD scan through the HMTP 0002 reflection is shown in Figure 6d features Laue oscillations, confirming a small amount of crystal faults and uniform layer thickness across the sample surface. The morphology of HMTP film on the quasi-freestanding SLG, measured with AFM, is shown in Figure 6b. It features a relatively flat film segmented by sub-μm- to several μm-long cracks. The RMS roughness of the film is ~7 nm, i.e., slightly larger than for SLG but smaller than for the buffer layer.

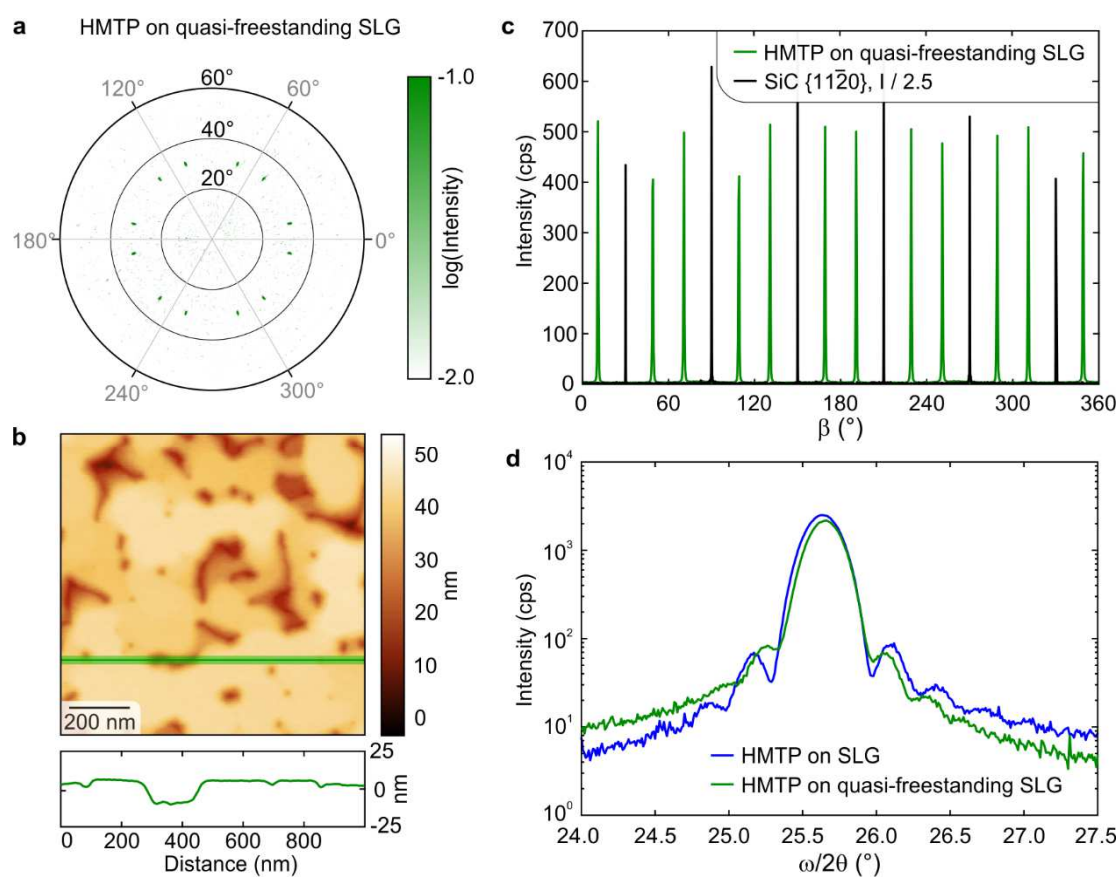

**Figure 6**: Analysis of HMTP thin film grown on quasi-freestanding SLG. (a) Pole figure, (b) morphology measured by AFM for a HMTP thin film grown on quasi-freestanding SLG, (c) azimuthal, and (d) symmetric ω/2θ scan measured through the HMTP 0002 reflection. Data in (a) and (c) were aligned to match their angular positions, differing due to the different placement of the sample in the diffractometer. The SiC intensity in (c) is downscaled by a factor of 2.5 for clarity.

**Conclusions**

In conclusion, we have used a combination of LEEM and XRD to demonstrate how crystalline order develops from the first layer to the thin film of an organic semiconductor, HMTP. We have employed two structurally similar substrates – single-layer graphene on SiC and its precursor, the buffer layer. On single-layer graphene, HMTP grows epitaxially: from the first layer, it forms large grains with precise alignment with the substrate. Conversely, on the buffer layer, growth is initially amorphous, while the HMTP molecules maintain their coplanar orientation with respect to the substrate throughout the growth. In the thicker layer, a crystalline order develops; however, the grains are small, display weak preferential orientation with respect to the substrate, and have non-uniform thickness. Our multiscale approach allows us to trace the in-plane polycrystallinity to hindered diffusion of HMTP on the non-uniform buffer layer, which shows a significant fraction of $sp^3$-hybridized carbon atoms that mediate bonding with the SiC substrate. Removing these bonds via hydrogen passivation of the SiC substrate results in the formation of a quasi-free-standing monolayer graphene, on which HMTP exhibits epitaxial growth. Interface engineering via hydrogen intercalation thus enables control over HMTP epitaxy on graphene.

**Methods**

**Sample Preparation.**

**Substrate preparation.** Epitaxial graphene was synthesized via the thermal decomposition of SiC in a 1 atm argon atmosphere. Commercial 6H-SiC:V semi-insulating wafers (Coherent/II-VI Inc.) were diced into 5×5 mm² samples and subjected to a standard cleaning procedure using acetone, isopropanol, and distilled water. The growth process was performed in a furnace, where the buffer layer was first formed at 1600 °C for 5 minutes. Single-layer graphene (SLG) was obtained by increasing the growth temperature to 1650 °C for 5 minutes, and quasi-free-standing SLG was fabricated through hydrogen intercalation of the buffer layer (1120°C), following the SLG[39,40] and quasi-free-standing SLG[41] growth protocols established in our previous work.

**HMTP deposition.** HMTP thin films with a nominal thickness of 28–33 nm were deposited from powder (Merck) by an effusion cell (MBE Komponenten OEZ) using a resistively heated quartz crucible operated at 175 °C in a deposition chamber with a base pressure of 5 ×10⁻¹⁰ mbar. Prior to deposition, the molecules were thoroughly degassed, and all substrates were annealed under ultra-high vacuum at 550 °C for 10 minutes to remove surface contaminants. During the deposition, the substrates were held at room temperature, i.e., below 30 °C. Identical deposition times of 60 minutes with the deposition rate of ~5 Å/min were used for all substrates, and the final film thickness was determined from X-ray Laue oscillations.

For LEEM experiments, the in situ, real-time growth of HMTP was investigated up to a nominal thickness of ~3.5 nm. HMTP molecules were deposited from an organic material evaporator (MBE Komponenten, Quad Cell OEZ40) using a resistively heated quartz crucible operated at 170 °C, with a deposition rate of ~0.5 Å/min, i.e., one order of magnitude lower than that used for thick film growth due to chamber design.

**Sample Characterization.**

**X-ray Diffraction**. XRD measurements were carried out using a Rigaku SmartLab 3 X-ray diffractometer equipped with a rotating Cu anode (wavelength of 0.154 nm) and a five-circle goniometer. For all XRD measurements, the incident X-ray beam was collimated and monochromatized by a parabolic multilayer mirror. For azimuthal, symmetric and rocking scans, the incident beam size was defined by a vertical and a horizontal slit; the vertical angular acceptance at the detector side (i.e., for the scattered beam) was determined by a pair of vertical receiving slits – the first positioned after the sample and the second in front of the detector. The horizontal collimation was realized by Soller slits, and the diffracted signal was detected either by a scintillation detector or a one-dimensional detector (D/teX Ultra) set to integration mode. For azimuthal scans, the openings of the vertical and horizontal incidence slits were 0.5 and 5 mm, respectively. The heights of the receiving slits were 2 mm, and 4 mm and the horizontal collimation was realized by a pair of 0.5° Soller slits – one in the incidence beam and the second at the detector arm, which defined the azimuthal resolution of the scan. For symmetric and rocking curve scans, a two-bounce Ge(220) channel-cut monochromator was inserted in the incident beam path to further reduce the beam divergence to 0.02° and to select only the Cu K$\alpha_1$ radiation. The incident vertical and horizontal slits of the widths of 0.2 mm and 2 mm, respectively, were used together with a 5° Soller slit at the detector side. For rocking curve measurements, an additional two-bounce Ge(220) channel-cut analyzer monochromator was placed at the detector arm in front of the Soller slit to achieve an angular resolution better than 0.01°. For pole figure measurements, a pinhole and a collimator with diameters of 0.3 mm and 0.2 mm, respectively, were used to collimate the incident beam and to reduce its size. The diffracted signal was collected using a two-dimensional detector (HyPix-3000) positioned 150 mm from the sample, enabling the simultaneous acquisition of data over multiple scattering and polar angles. Subsequently, pole figures for required reflections were extracted from a series of 2D images using the 2DP software by Rigaku Holdings Corp.

**Atomic Force Microscopy (AFM).** AFM measurements were performed in tapping mode using a Bruker Dimension Icon microscope, and the AFM data were processed using Gwyddion software.[42]

**Raman spectroscopy.** The epitaxial graphene samples were characterized using a WITec alpha300 RSA confocal micro-Raman system. Measurements were performed in a backscattering geometry using a 532 nm excitation laser with a power of 20 mW. The signal was collected through a Zeiss microscope objective (100× magnification, NA = 0.9). To ensure high spatial resolution and signal-to-noise ratio, the system uses an optical fiber as a confocal pinhole to couple light from the microscope to the spectrometer.

**Low-Energy Electron Microscopy and Diffraction (LEEM/LEED)** experiments were performed using a Specs FE-LEEM P90 system operated at a base pressure of approximately $2\times10^{-10}$ mbar. Samples introduced into the LEEM system from ambient conditions were annealed at 550 °C for 60 min prior to measurements. Bright-field LEEM images were obtained by collecting electrons from the specular (0,0) beam. LEED patterns were recorded from surface areas of 15×10 $\mu m^2$, while single-domain diffraction measurements were carried out using an electron beam with a diameter of 185 nm. Diffraction data were analyzed using ProLEED Studio.[43]

ASSOCIATED CONTENT

Supplemental Material: (1) Raman Spectroscopy characterization of SLG and buffer layer substrates; (2) XRD  and AFM measured on samples with a coexisting buffer layer and SLG; (3) Full-scale symmetric scan; (4) The large-scale AFM images measured on the buffer and SLG; (5) LEED model of graphene on SiC and moiré structure; (6) LEED model of HMTP on graphene; (7) LEEM dark-field analysis of HMTP on SLG; (8) LEED single-domain diffraction of HMTP on the buffer layer.

AUTHOR INFORMATION

**Corresponding Authors**

*E-mails:    jan.kunc@matfyz.cuni.cz    (J.K.);    novak@physics.muni.cz    (J.N.); cechal@fme.vutbr.cz (J.Č.)

**Author Contributions**

D.V. prepared the samples for ex-situ measurements and performed and evaluated AFM measurements; D.V. and J.N. measured and evaluated the XRD data; P.P. together with V.S. measured and evaluated the LEEM/LEED data; M.S. and J.K. prepared the substrates; J.Č., J.K., and J.N. conceived, coordinated and supervised the project; J.Č. wrote the manuscript through the contribution of all authors; all authors discussed the results and approved the final manuscript.

COMPETING INTERESTS

The authors declare no competing interests.

ACKNOWLEDGMENT

This research has been supported by GAČR, project No. 22-04554S. D.V. and J.N. additionally acknowledge the support by the project Quantum Materials for Applications in Sustainable Technologies, Grant No. CZ.02.01.01/00/22_008/0004572. We acknowledge CzechNanoLab Research Infrastructure (LM2023051) supported by MEYS CR.

DATA AVAILABILITY

Data for this article, including XRD, LEEM, LEED, and AFM images, and employed scripts are available at Zenodo at [the link will be provided at the revision step].

# Interfacial Coupling Controls Molecular Epitaxy of HMTP on Graphene/SiC

*Devanshu Varshney,[1&] Pavel Procházka,[2&] Veronika Stará,[2] Mykhailo Shestopalov,[3] Jan Kunc,[3*] Jiří Novák,[1*] Jan Čechal[2,4*]*

[1]Department of Condensed Matter Physics, Faculty of Science, Masaryk University, Kotlářská 2, 61137 Brno, Czech Republic

[2]CEITEC - Central European Institute of Technology, Brno University of Technology, Purkyňova 123, 612 00 Brno, Czech Republic.

[3] Charles University, Faculty of Mathematics and Physics, Institute of Physics, Ke Karlovu 5, 121 16, Prague 2, Czech Republic

[4] Institute of Physical Engineering, Brno University of Technology, Technická 2896/2, 616 69 Brno, Czech Republic.

[&]These authors contributed equally.

CONTENTS:

# 1. Raman Spectroscopy characterization of SLG and buffer layer substrates

As detailed below, a typical SLG shows a Lorentzian 2D peak with a FWHM of 32–40 cm$^{-1}$ and an integrated 2D-to-G peak intensity ratio of 1.8 ± 0.3, typical fingerprints of SLG. The quasi-freestanding SLG shows a narrower 2D peak with FWHM of 23–33 cm$^{-1}$, and patches of bilayer graphene with FWHM in the 50–56 cm$^{-1}$ range.

## 1A. Comparison of the SLG and quasi-freestanding SLG samples:

The ratio of G to D peak integrated intensities (Figure S1a) is proportional to the graphene grain size. It is ~2 for SLG because the underlying buffer layer contributes to the enhanced D peak intensity. The G to D peak ratio approaches ~5 for quasi-freestanding SLG, which corresponds to the grain size of about 100 nm when using 532 nm laser excitation.[1] The ratios of integrated 2D to G peak intensity of 1.8 ± 0.3 and 4.5 ± 0.5 given in Figure S1b are characteristic to SLG and quasi-freestanding SLG, respectively. The strain given in Figure S1c was determined from the G and 2D peak positions. The quasi-freestanding SLG shows a partially relaxed strain compared to SLG samples.

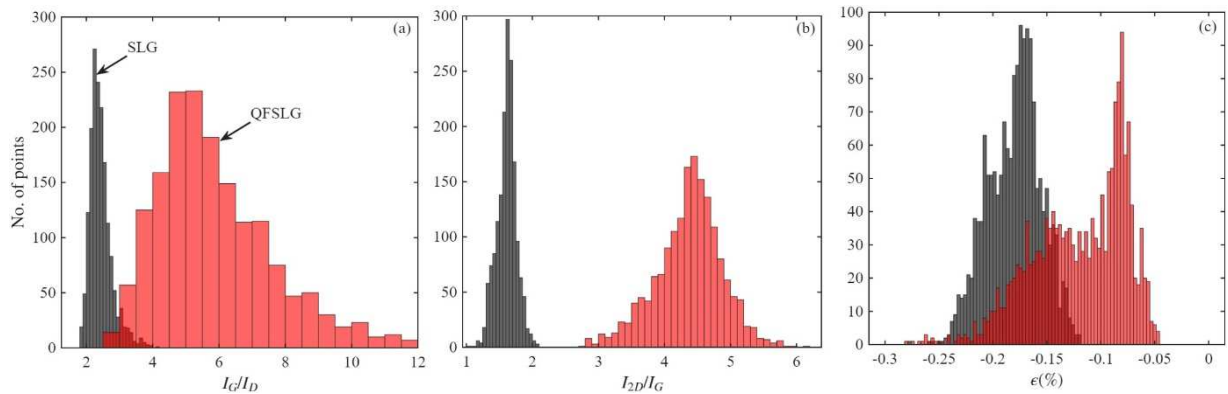

**Figure S1**: Comparison of Raman peak intensities measured on the SLG and quasi-freestanding SLG (QFSLG) samples: (a) G-to-D peak ratio and (b) 2D-to-G peak ratio. (c) Calculated elastic strain in the layer.

The 2D peak position is governed mostly by mechanical strain. The redshifted 2D peak in Figure S2a reflects a partially relaxed strain of the quasi-freestanding SLG. The improved quality of quasi-freestanding SLG compared to SLG is also revealed by the improved FWHM of 2D peak from $33 \pm 2$ cm$^{-1}$ for SLG to $23 \pm 2$ cm$^{-1}$ for quasi-freestanding SLG (Figure S2b). The FWHM of the 2D peak is a fingerprint of inhomogeneous strain on a sub-micrometer length scale.

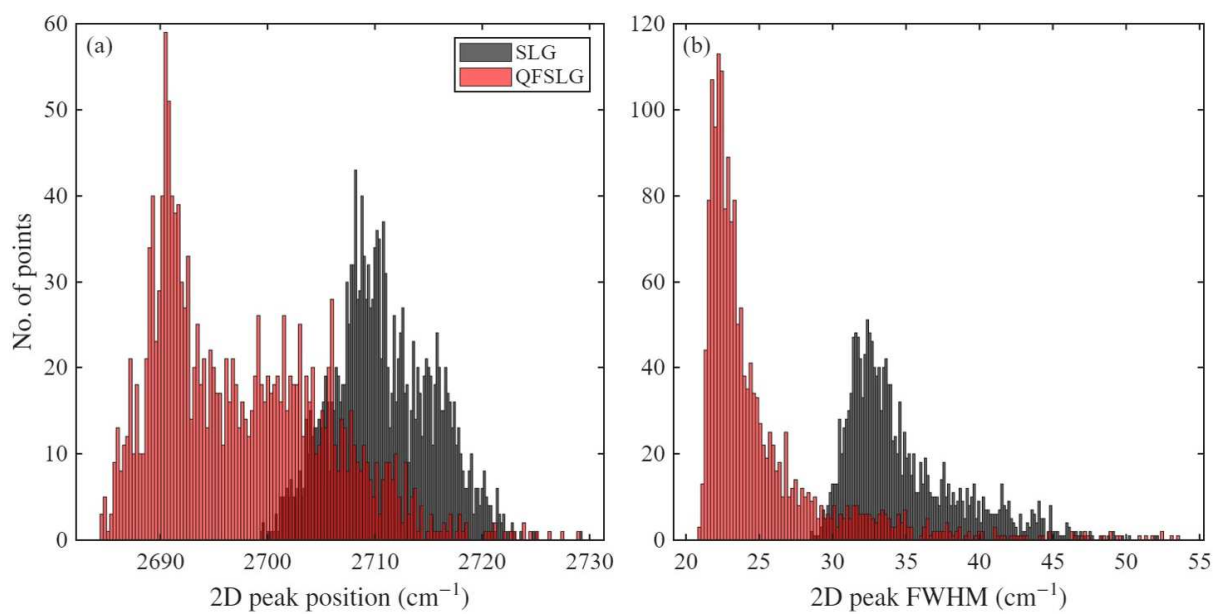

**Figure S2**: Comparison of Raman 2D peak position and FWHM measured on the SLG and quasi-freestanding SLG (QFSLG) samples: (a) 2D peak position and (b) 2D peak FWHM.

## 1B. Comparison of buffer layer samples

The buffer layer before the hydrogen intercalation is the precursor used to fabricate the quasi-freestanding SLG sample. A small ratio of the G to D peak integrated intensities $1.2 \pm 0.2$ (Figure S3a) reveals a small grain size of the buffer layer around $25 \pm 5$ nm. The broad G peak in Figure S3b indicates a largely distorted graphene-like lattice. The nearly zero 2D peak intensity relative to the G peak intensity (Figure S3c) indicates an undeveloped graphene electronic band structure.

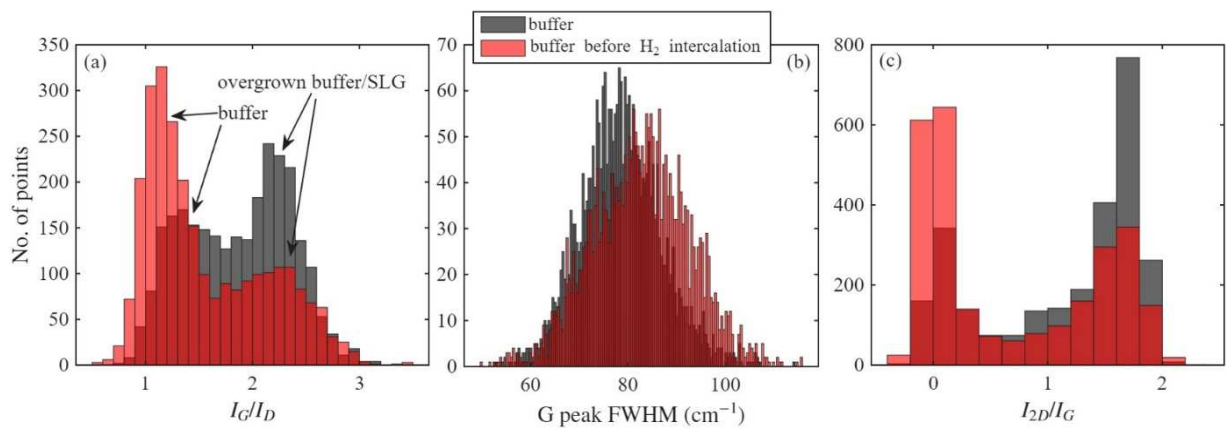

**Figure S3**: Comparison of Raman peak signals measured on the two distinct samples with the buffer layer: one was used for HMTP deposition, and the second for preparation of quasi-freestanding SLG; this shows a sample-to-sample variability of buffer layers. (a) G-to-D peak ratio, (b) FWHM of the G peak, and (c) 2D-to-G peak ratio.

## 1C Sample homogeneity and typical Raman spectra

To assess the homogeneity of the SLG samples, Raman maps were clustered into five areas and further reduced to typical graphene allotropes, as shown in Figures S4a, d, and g. The corresponding 2D peak spectra are shown in Figures S4b, e, and h, and the number of 2D peak components can be determined from the number of minima in the second derivative given in Figures S4c, f, and i.

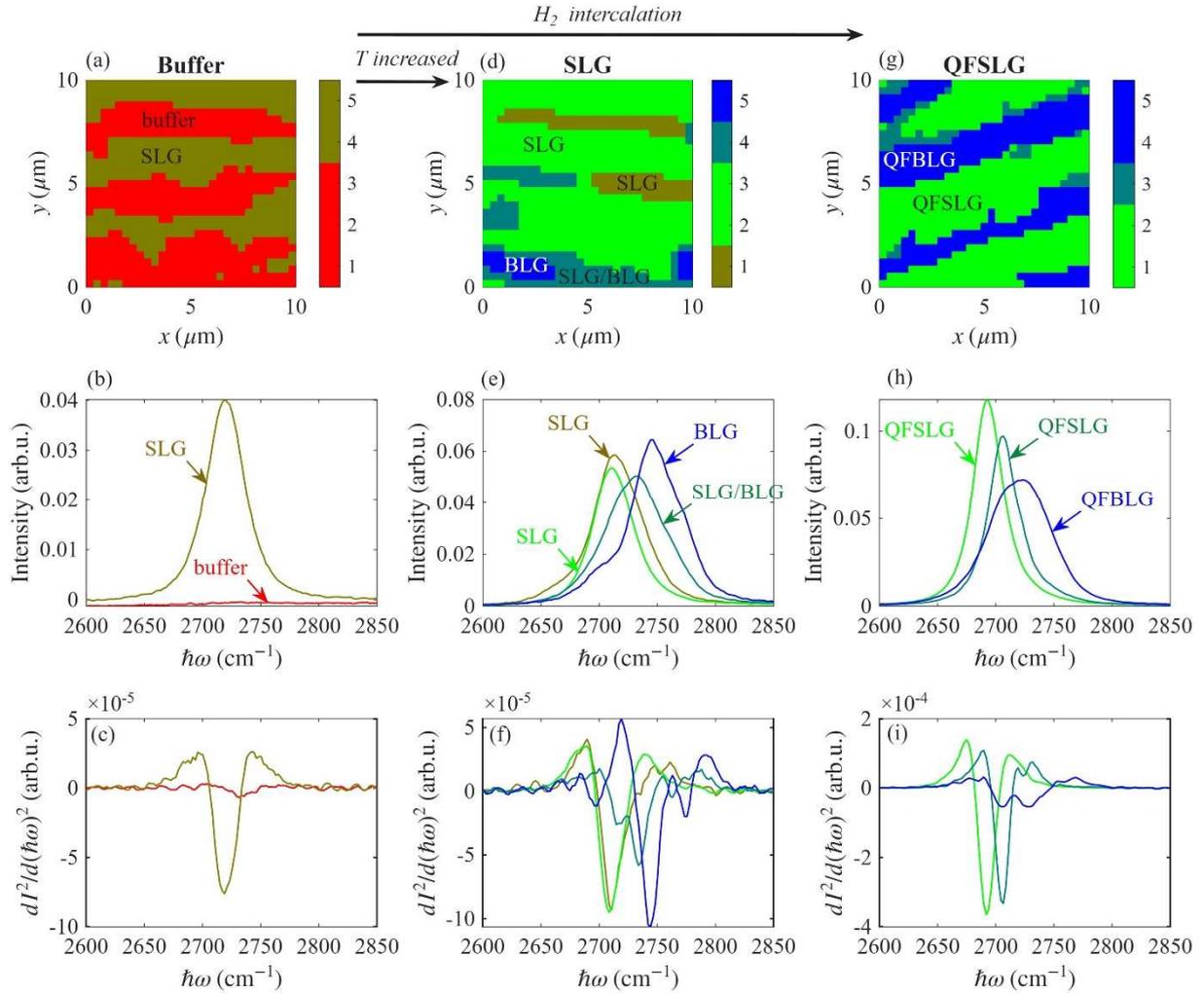

**Figure S4**: Raman maps clustered into five different areas on (a) the buffer layer, (d) SLG, and (g) quasi-freestanding SLG (QFSLG). The clusters are color-coded for (red) buffer, (dark yellow) strained SLG or mixture of SLG and bilayer graphene, (light green) SLG in (d) and QFSLG in (g), (dark cyan) transition between SLG and bilayer graphene (BLG) and QFSLG to BFLG (in (d) and (g), respectively), (blue) BLG and QFBLG (in (d) and (g), respectively). Raman spectra of the 2D peak region associated with clusters in (a, d, and g) measured on the (b) buffer, (e) SLG, and (h) quasi-freestanding SLG (QFSLG). The spectra are averaged within each of the five clusters. The second derivative (c, f, and i) shows the number of 2D peak components. A single-component 2D peak represents SLG; the four-component 2D peak represents bilayer graphene. Bilayer graphene typically grows at SiC step edges.

## 2. XRD and AFM measured on samples with a coexisting buffer layer and SLG

The diffraction data for the substrate with a coexisting buffer layer and SLG (approximate 2:1 coverage ratio), shown in Figure S5, correspond to the sum of the diffraction patterns from samples with only the buffer layer and only SLG, respectively. The pole figure of HMTP $\{10\bar{1}1\}$ shows two sets of six sharp spots at radial and azimuthal positions identical to the SLG sample (Figure 2 in the main text), together with a slightly enhanced background. The orientational order characteristic of the HMTP on SLG dominates the scattering pattern, despite the buffer layer covering most of the substrate. The detailed analysis of the in-plane directional ordering was done by azimuthal scans probing HMTP $\{10\bar{1}1\}$ planes radially tilted out of surface plane by 31.7° (Figures S5b and c). Also, here, we observe sharp and intense HMTP $\{10\bar{1}1\}$ reflections offset by ±19.1° out of nearest SiC $\{11\bar{2}0\}$ reflections, corresponding to the intense spots in the pole figure, and indicating the same rotation angle of the HMTP lattice as observed for the SLG sample. Additionally, there are less intense and broader peaks corresponding to HMTP on the buffer layer, where HMTP $\{10\bar{1}1\}$ reflections are offset by ±11° out of nearest SiC $\{11\bar{2}0\}$ reflections. Their intensity is suppressed compared to the corresponding peaks on the buffer sample (see Figure 2 in the main text), and their maximum is shifted with respect to the pure buffer sample (the offset of ±8.5°). This suggests that the preferential in-plane rotation of the HMTP domains on the buffer is driven by the amount of contact with SLG domains with which they coexist.

The AFM images of the HMTP film grown on the sample with coexisting buffer and SLG are shown in Figure S6. A relatively flat percolated HMTP film decorates the SLG part that dendritically grows from SiC surface step edges, while small islands cover the buffer, with diameters in the hundreds of nm. This picture is consistent with AFM images of the SLG and buffer samples (Figure 4 in the main text and Figure S8), thus highlighting the coexistence of the two types of HMTP film morphologies next to each other on the SiC substrate with coexisting buffer and SLG areas.

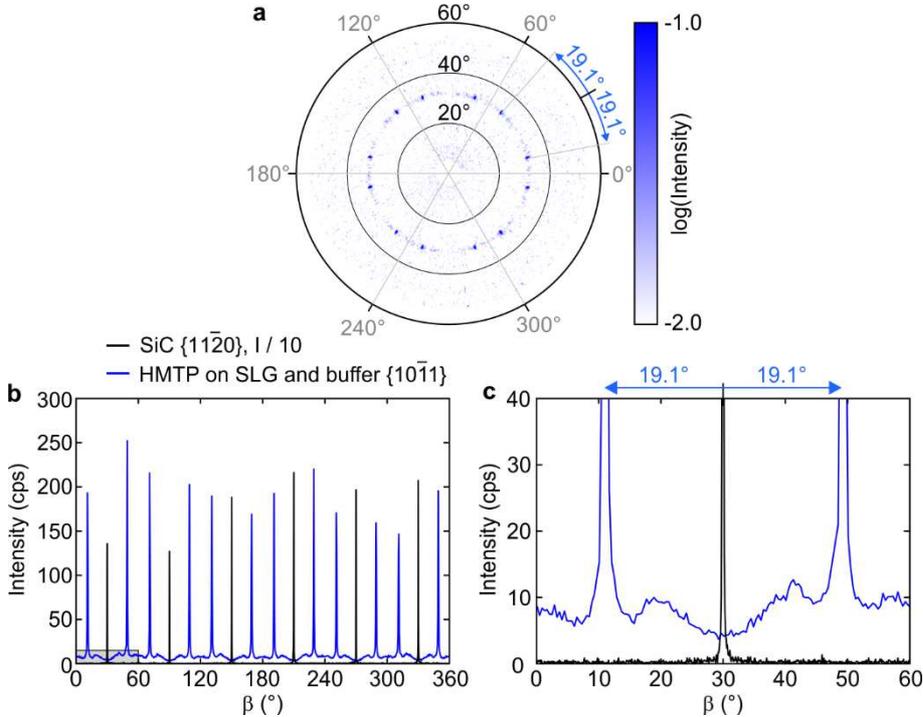

**Figure S5**: (a) Pole figure of HMTP {10$\bar{1}$1} reflections and (b) azimuthal (β) scan of the HMTP {10$\bar{1}$1} reflections at the polar angle 31.7° for an HMTP film on a substrate with a coexisting buffer layer and SLG (approximate 2:1 coverage ratio). In the azimuthal scans, the data for HMTP {10$\bar{1}$1} (blue) are shown together with SiC substrate {11$\bar{2}$0} in-plane reflections (black). The azimuthal angle β was aligned using the SiC {11$\bar{2}$0} reflections as reference. The SiC intensity is downscaled by a factor of 10 for clarity. (c) An enlarged view of the azimuthal scans.

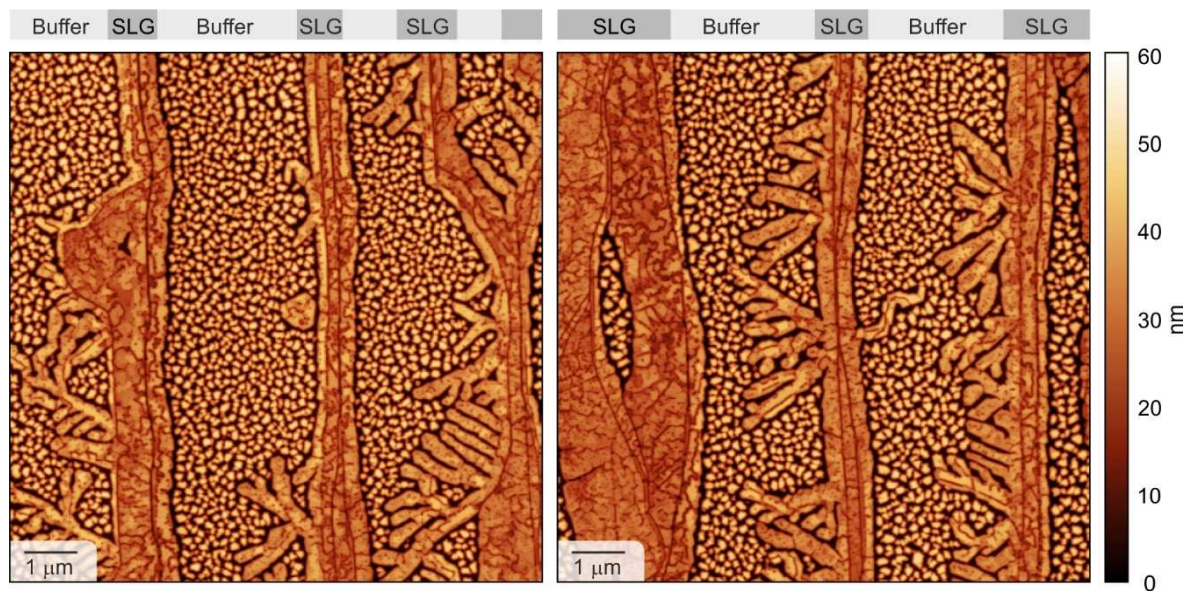

**Figure S6**: Large-scale AFM images of HMTP on a sample with coexisting SLG and the buffer layer (estimated coverage ratio 1:2). Identification of the respective areas is given on top of the figure.

## 3. Full-scale symmetric scan

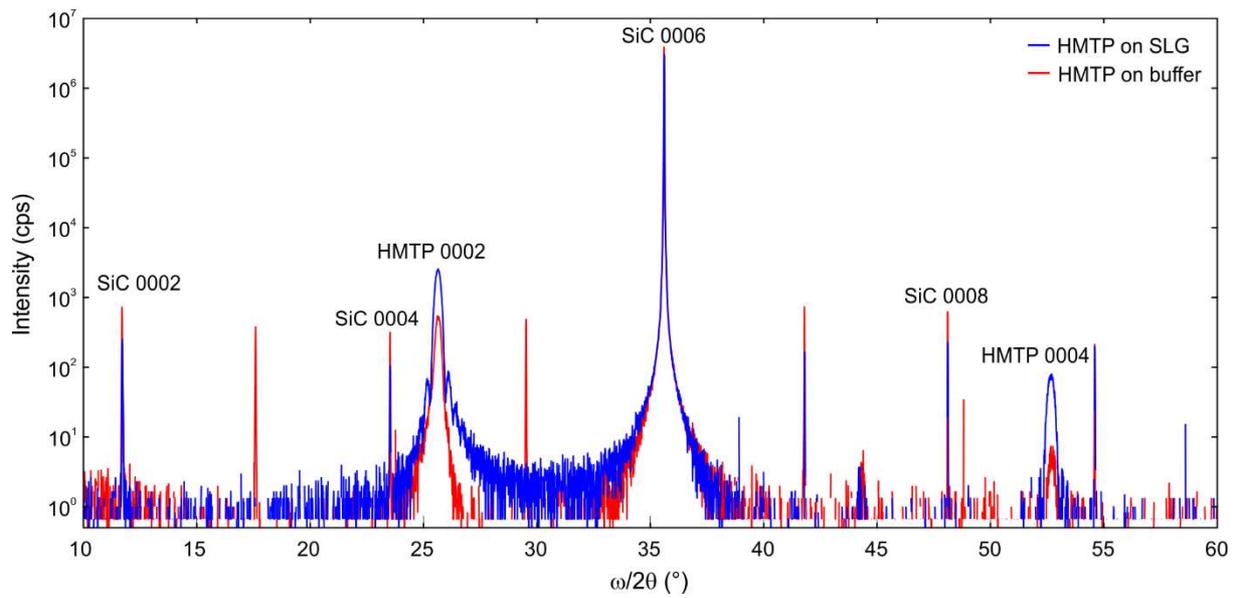

**Figure S7:** Overview of a symmetric ω/2θ scan for HMTP thin film on SLG (red) and the buffer layer (black). The marked peaks correspond to HMTP 000*l* and SiC 000*l* reflections from Cu Kα₁ radiation. The unmarked intense peaks originate from SiC 000*l* diffraction of higher harmonics.

## 4. The large-scale AFM images measured on the buffer and SLG

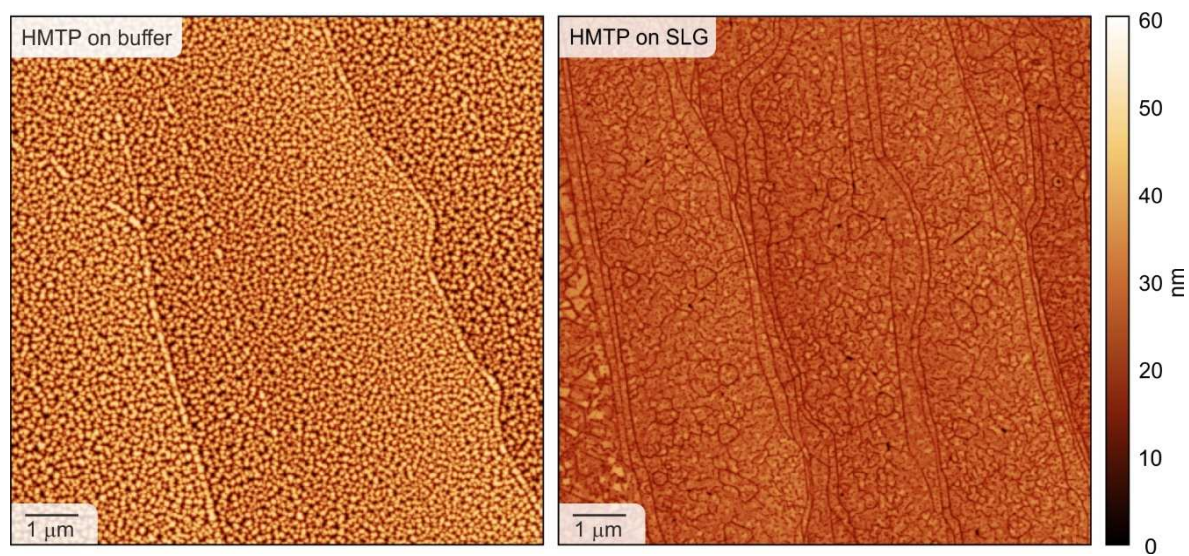

**Figure S8**: Large-scale AFM images of the HMTP film on the buffer layer and SLG associated with Figure 4 in the main text.

## 5. LEED model of graphene on SiC and moiré structure

Figure S9a illustrates the diffraction model of epitaxial graphene on SiC used to interpret the LEED features discussed in the main text. The diffraction pattern (Figure S9b) measured at 60 eV consists of graphene spots, SiC substrate spots, and additional satellite (moiré) spots arising from their relative lattice mismatch and rotational alignment. The corresponding model, generated in ProLEED Studio,[2] overlays the reciprocal lattices of graphene and SiC and reproduces the positions of the observed moiré spots.

To visualize the structural origin of these features, the real-space SiC and graphene lattices are shown together (Figure S9a) with their respective unit cells. The large-scale moiré superlattice emerges from the mismatch between the 2.46 Å graphene lattice and the 3.08 Å SiC lattice and corresponds to a $(6\sqrt{3} \times 6\sqrt{3})R30°$ reconstruction with a periodicity of approximately 32 Å, as shown in Figure S9c,d. An enlarged view of the moiré diffraction in Figure S9c shows the moiré unit cell and the reciprocal-lattice vectors responsible for the satellite spots observed in LEED. In the main text, only the large, higher-order moiré spots are visible, while the primary moiré spots remain too weak to be detected.

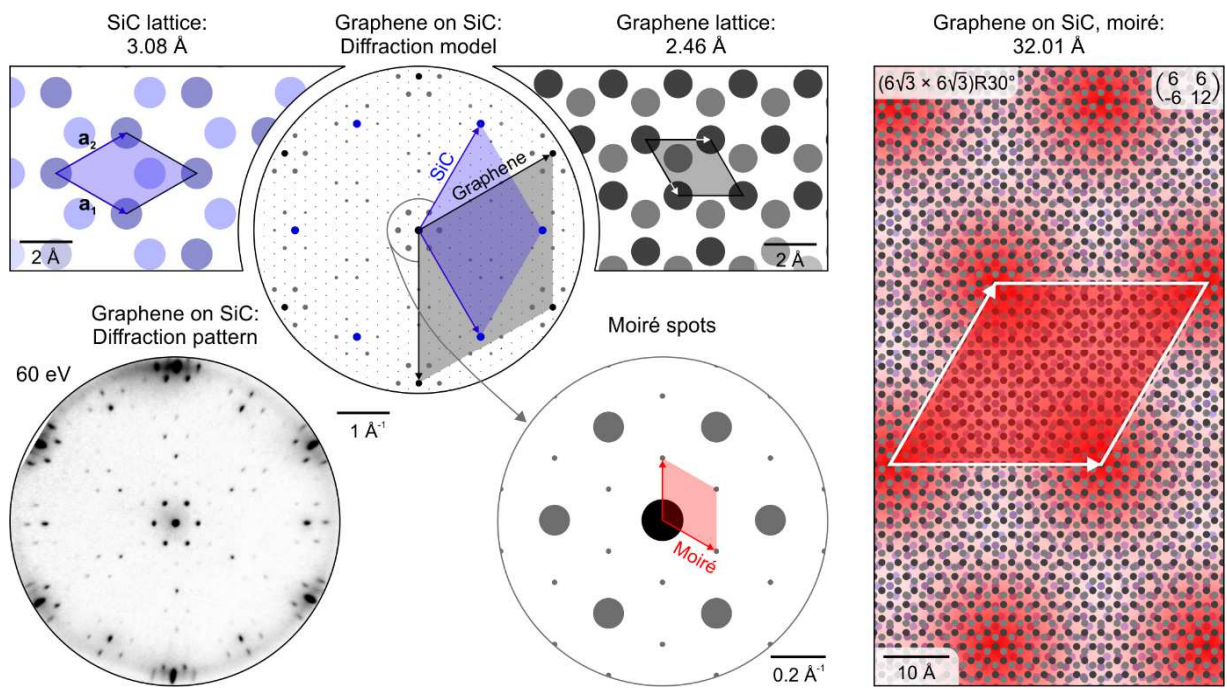

**Figure S9**: (a) Real-space SiC (3.08 Å) and graphene (2.46 Å) lattices with their unit cells. The ProLEED Studio diffraction model of graphene on SiC (center) shows the corresponding reciprocal lattice points of graphene (black) and SiC (blue). (b) Experimental LEED pattern measured at 60 eV. (c) Enlarged view of the diffraction model highlighting the moiré spots, with the moiré unit cell marked. (d) Real-space representation of the $(6\sqrt{3} \times 6\sqrt{3})R30°$ moiré superlattice with a periodicity of ~32 Å. The moiré spots originate from the relative lattice mismatch and 30° rotation between graphene and the SiC substrate.

## 6. LEED model of HMTP on graphene

Figure S10 summarizes the diffraction model used to interpret the HMTP overlayer on SLG. The full diffraction pattern in Figure S10a shows the combined contributions of the two symmetry-equivalent HMTP domains. A comparison with the simulated pattern confirms the assignment of the main diffraction spots. Single-domain diffraction patterns extracted from individual regions of the sample are shown in Figure S10b. Their corresponding simulated patterns in Figure S10c reproduce the positions of the domain-specific spots and verify that both domains adopt the same commensurate superstructure with different in-plane orientations.

Figure S10d presents the real-space structural model of the HMTP overlayer with respect to the graphene lattice for each domain. The superstructure corresponds to the $\begin{pmatrix} 4 & 2 \\ 6 & -4 \end{pmatrix}$ matrix notation and describes a $2\sqrt{7} \times 2\sqrt{7}$ R19.1° commensurate unit cell. The two equivalent HMTP orientations differ only by their rotational alignment relative to graphene.

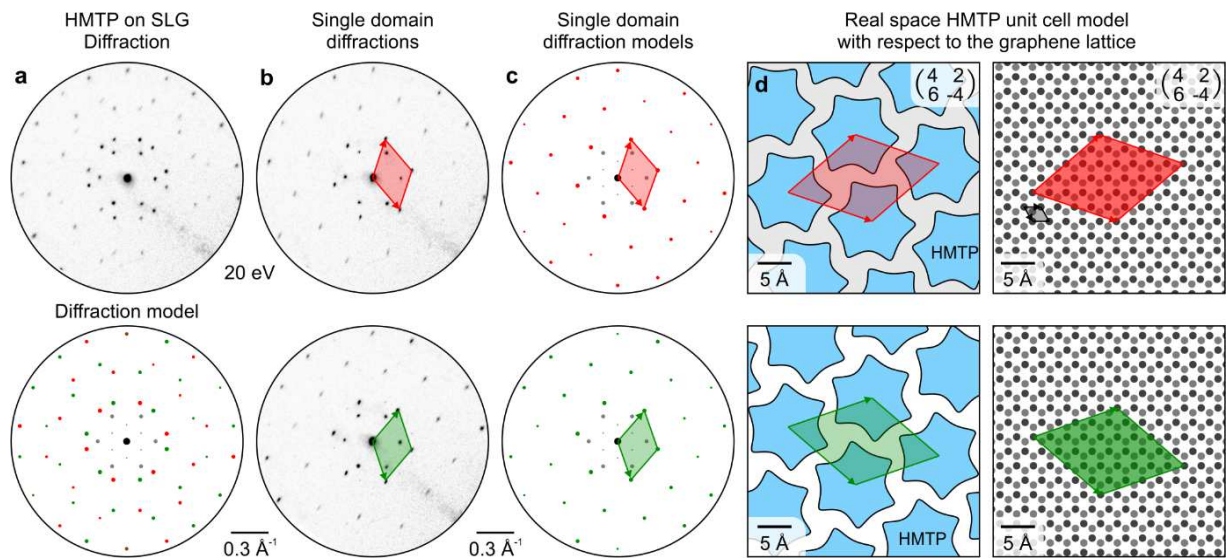

**Figure S10**: HMTP diffraction model on SLG. (a) Experimental HMTP diffraction pattern on SLG at 20 eV together with its simulated model. (b) Single-domain diffraction patterns extracted from two HMTP domains. (c) Corresponding simulated patterns for the two domain

orientations. (d) Real-space structural model of the HMTP unit cell relative to the graphene lattice for both orientations, described by the $\begin{pmatrix} 4 & 2 \\ 6 & -4 \end{pmatrix}$ matrix notation corresponding to a $2\sqrt{7} \times 2\sqrt{7}$ R19.1° superstructure.

## 7. LEEM dark-field analysis of HMTP on SLG

Figure S11 shows the individual dark-field images used to construct the composite DF presented in Figure 5f of the main text. Each dark-field image was obtained by selecting one of the two HMTP diffraction spots highlighted in the LEED pattern, isolating a single HMTP domain orientation on SLG. The resulting images, together with their color-coded composite, confirm that ordered HMTP domains form exclusively on SLG, while no crystalline contrast is detected on the buffer layer.

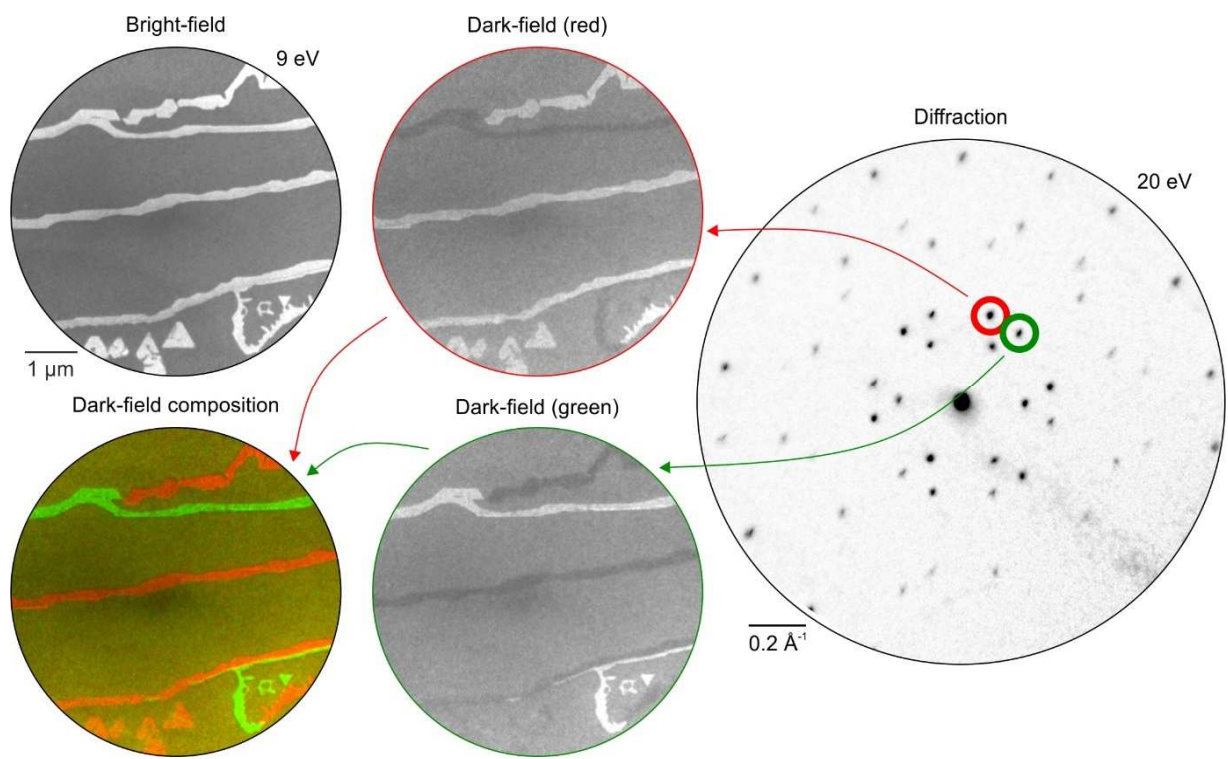

**Figure S11**: Individual dark-field images underlying the composite shown in Figure 5f of the main text. The figure includes the bright-field image (9 eV), the LEED pattern (20 eV) with the selected diffraction spots, and the corresponding dark-field images for the two HMTP orientations on SLG. The color-coded composite maps the spatial distribution of both orientations. Ordered HMTP is observed only on SLG, with no crystalline contrast on the buffer layer.

## 8. LEED single-domain diffraction of HMTP on the buffer layer

As shown in Figure S12, single-domain LEED patterns measured on the buffer layer display only the characteristic moiré spots of graphene/SiC. No additional diffraction spots associated with HMTP are observed, in contrast to the ordered domains on SLG. This confirms that HMTP does not form a crystalline overlayer on the buffer layer.

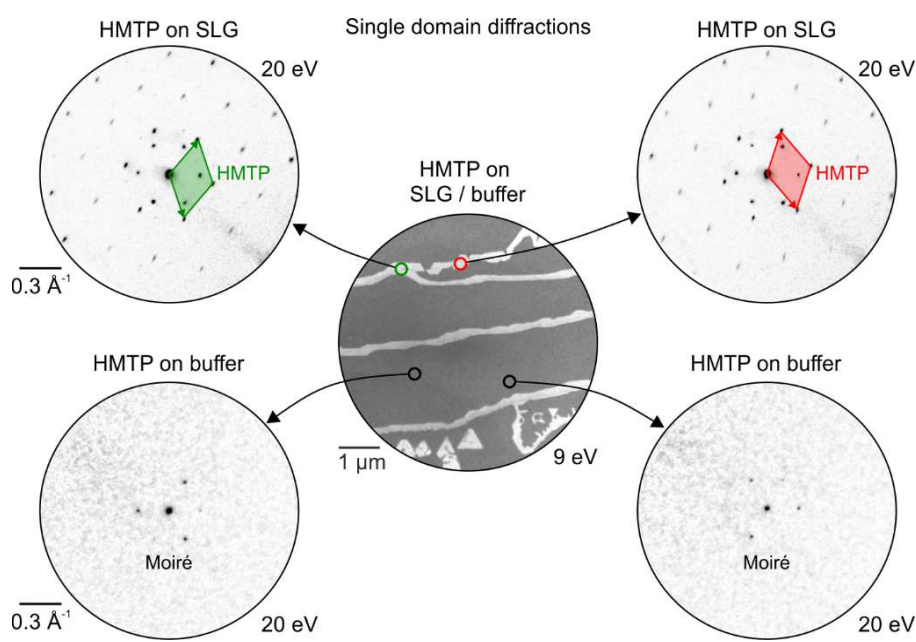

**Figure S12**: Single-domain diffraction measurements of HMTP on SLG and the buffer layer. The bright-field image (center) indicates the probed positions. On SLG (top), two distinct HMTP domain orientations produce sharp diffraction spots. On the buffer layer (bottom), only the substrate-related moiré spots are present, with no additional diffraction attributable to HMTP.